\documentclass[aps,preprint]{revtex4}%
\usepackage{amsfonts}
\usepackage{amsmath}
\usepackage{amssymb}
\usepackage{graphicx}%
\providecommand{\U}[1]{\protect\rule{.1in}{.1in}}

\begin{document}
\title[ ]{Stochastic Electrodynamics: The Closest Classical Approximation to Quantum Theory}
\author{Timothy H. Boyer}
\affiliation{Department of Physics, City College of the City University of New York, New
York, New York 10031}

\begin{abstract}
Stochastic electrodynamics is the classical electrodynamic theory of
interacting point charges which includes random classical radiation with a
Lorentz-invariant spectrum whose scale is set by Planck's constant. \ Here we
give a cursory overview of the basic ideas of stochastic electrodynamics, of
the successes of the theory, and of its connections to quantum theory. \ 

\end{abstract}
\maketitle

\section{Introduction}

\subsection{Two Regimes: Classical and Quantum}

Physical theory operates in two different regimes. \ The familiar macroscopic
word is classical, dealing with waves and particles which experience friction
and can be described in terms of continuously changing forces, energies, and
angular momentum. \ On the other hand, the microscopic world of atomic and
molecular physics is usually described quantum mechanically with intrinsic
randomness, wave-particle duality, entanglement, and discrete values for
energy and angular momentum. \ Occasionally frictionless phenomena
(appropriate for the quantum world) intrude into the classical macroscopic
world in the form of persistent currents in superconducting wires or
superfluids which flow over gravitational potential barriers. \ Furthermore,
phenomena such as the blackbody radiation spectrum and the decrease of
specific heats at low temperatures do not seem to follow classical
descriptions given by traditional classical statistical mechanics. \ The
currently-accepted descriptions of these intruding phenomena are traced back
to quantum behavior for the constituents at the microscopic level.
\ Accordingly, the question arises as to the location of the \textquotedblleft
microscopic\textquotedblright\ level. \ At what level of interaction for
energy or angular momentum are continuous ideas of classical physics no longer
applicable? \ Where is the quantum-classical boundary? \ 

\subsection{Boundary Based Upon Planck's Constant}

According to the elementary theory taught commonly at present,\cite{ER} the
boundary between classical and quantum physics is given by the appearance of
Planck's constant $h$ or $\hbar=h/(2\pi).$ \ If the action variable $J$ for a
system is of the order of $\hbar$, then the system must be described within
quantum theory. \ Thus the angular momentum of an atom or molecule must come
in discrete values involving $\hbar.$ \ If an oxygen molecule has a
vibrational energy $\mathcal{E}=\omega\,J$, then the energy must come in
discrete multiples $J=\left(  n+1/2\right)  \hbar,$ $n=0,1,2,...$ \ In
instructional university physics, the transition between classical and quantum
physics is given by the presence Planck's constant: Planck's constant does not
not appear within traditional classical physics, and Planck's constant is the
crucial constant entering the quantum domain.\cite{deco}\ 

\subsection{Stochastic Electrodynamics Reopens the Boundary Question}

Stochastic electrodynamics (SED) is a theory of classical electrodynamics
where the homogeneous (source-free) boundary condition on Maxwell's equations
is taken to include random classical radiation with a Lorentz-invariant
spectrum which is supposed to exist in all spacetime.\cite{M}\cite{B1969}%
\cite{PC} \ The scale of this Lorentz-invariant random radiation is set by
Planck's constant $\hbar$. \ Thus stochastic electrodynamics is a classical
theory which incorporates Planck's constant. \ However, if we allow stochastic
electrodynamics as a theory of classical physics, then we have Planck's
constant included within both classical and quantum theories. \ Clearly, the
traditional classical-quantum distinction favored within main-steam academic
circles is no longer valid. \ Indeed, the classical theory of stochastic
electrodynamics can describe accurately some of the phenomena (such as
blackbody radiation, the decrease of specific heats at low temperatures, and
the absence of atomic collapse) which are currently claimed to have
explanations only in terms of quantum theory. \ Thus the question of the
boundary between classical and quantum theories has to be reconsidered, and
not just in some sophisticated \textquotedblleft quantum\textquotedblright%
\ effect of decoherence. \ Stochastic electrodynamics (which contains Planck's
constant) goes beyond the results of traditional classical theory (which does
not include Planck's constant) and comes closer to giving the results of
quantum theory. \ Indeed, stochastic electrodynamics is at present the closest
classical approximation to quantum theory. \ The unanswered question
remains:\ \textquotedblleft How many of the phenomena which are currently
regarded as requiring quantum explanations can actually be described by the
classical theory of stochastic electrodynamics?\textquotedblright\ \ 

\section{Foundations of Stochastic Electrodynamics}

\subsection{Theoretical Assumptions}

Stochastic electrodynamics is a relativistic theory of point charges and
electromagnetic fields based on three fundamental ideas. \ 1) The
electromagnetic fields satisfy Maxwell's equations with sources given by point
charged particles. \ 2) The particles experience forces due to electromagnetic
fields and move according to Newton's second law with the force given by the
Lorentz force law, $d\mathbf{p}/dt=e\mathbf{E}+e\mathbf{v\times B/c.}$ \ 3)
The solution of the source-free Maxwell equations is given by classical
electromagnetic zero-point radiation; namely the Lorentz-invariant spectrum of
random classical radiation with an energy per normal mode $\mathcal{E(}%
\omega)=\hbar\omega/2.$ \ The basis for these assumptions comes from
experimental work in macroscopic electrodynamics (which forms the foundations
for much of our electrical technology) together with the inference of random
classical zero-point radiation which arises from the experiments on Casimir
forces between conducting parallel plates.

\subsection{Inference from Measurements on the Casimir Effect}

The presence of classical electromagnetic zero-point radiation is inferred
from Casimir force measurements\cite{Lam} at low temperature. \ The Casimir
effect is a force between uncharged, parallel conducting plates of area $A$
and separation $d$. \ Within classical theory, the force between the plates
arises from the ambient radiation which surrounds the plates. \ At high
temperatures $T$, the (attractive) force $F$ between the plates goes as
$F=-[\zeta(3)/(4\pi)]k_{B}TA/d^{3},$ (where $k_{B}$ is Boltzmann's constant)
associated with the thermal radiation surrounding the plates corresponding to
the Rayleigh-Jeans spectrum of thermal radiation with an energy $\mathcal{E(}%
\omega)=k_{B}T$ per normal mode at high temperature. \ If continued to zero
temperature $T\rightarrow0,$ this formula predicts that the force between the
plates vanishes. \ However, experimental measurements of the Casimir force at
low temperatures indicate that the force between the plates does not vanish at
low temperature but rather goes over to the temperature-independent form
$F=-[const\times\pi^{2}c/120]A/d^{4}.$ \ The non-zero temperature-independent
force holding at low temperatures indicates the presence of random classical
radiation even at the absolute zero of temperature. \ Furthermore, the
$d^{-4}$-functional form for the force indicates that the spectrum of the
random radiation must be given by $\mathcal{E(}\omega)=const\times\omega.$
\ If we choose the value of the constant $const$ so as to fit the experimental
data, we find that the value corresponds to a familiar constant,
$const=\hbar/2.$ \ Thus the natural inference in classical physics is that at
low temperatures the world contains random classical radiation with an energy
spectrum $\mathcal{E(}\omega)=\hbar\omega/2$ per normal mode, corresponding to
classical electromagnetic zero-point radiation. \ 

\subsection{Change in the Classical Outlook Compared with Current Classical
Physics}

The foundational ideas of currently-accepted classical physics assume that as
the ambient temperature decreases toward absolute zero, all ambient radiation
vanishes. \ This is the same erroneous assumption as was made a century ago by
the physicists in the early years of the 20th century. \ H. A. Lorentz, who
developed traditional classical electron theory, makes explicit this
assumption of vanishing ambient random radiation at the absolute zero of
temperature.\cite{HAL} \ Indeed, this assumption seems so natural to most
physicists, that it is rarely questioned. Because traditional classical
physics lacks the idea of classical zero-point radiation, the description of
phenomena at the microscopic level presents an impossible challenge.
\ Historically, the challenge was met only by turning to entirely different
\textquotedblleft quantum\textquotedblright\ ideas. \ 

\ The presence of classical electromagnetic zero-point radiation will have a
dramatic effect for microscopic systems when the energy changes associated
with the effects of zero-point radiation are large compared with other
energies in the situation. \ On the other hand, for macroscopic phenomena
involving large electromagnetic fields, \ the presence of zero-point radiation
or thermal radiation can be ignored as insignificantly small. \ However, for
microscopic phenomena at the atomic scale, the zero point energy
$\mathcal{E(}\omega)=\hbar\omega/2$ can be comparable to the atomic energies
of the problem, and so becomes vitally important to an understanding of the
phenomena. \ Stochastic electrodynamics is a classical theory involving the
same continuous ideas on both the macroscopic and microscopic scales, and so
is completely different from a quantum theory with its discrete values for the
action variable, energy, and angular momentum. \ Stochastic electrodynamics
also involves a boundary between macroscopic phenomena (where the
contributions of zero-point radiation are generally so small as to be
irrelevant) and microscopic phenomena (where the contributions of zero-point
radiation may be vital). \ However, the boundary is clearly defined in terms
of energy scales with no change in concepts as one crosses the boundary. \ 

\section{Successes of Stochastic Electrodynamics}

\subsection{Charged Harmonic Oscillator in Zero-Point Radiation}

The spectrum of classical zero-point radiation can be inferred from the
Casimir effect. \ Indeed the presence of classical zero-point radiation
accounts completely for the experimentally-measured forces associated with the
Casimir effect. \ However, the presence of classical zero-point radiation will
influence any other classical system. \ The easiest system to consider is a
charged harmonic oscillator because the equations of motion are
linear.\cite{M}\cite{B1969}\cite{PC}

A small dipole oscillator which is immersed in random classical radiation will
have an equation of motion given by $m\ddot{x}=-kx+m\tau\dddot{x}%
+eE_{x}(0,t),$ where the particle of charge $e$ and mass $m$ is attached to a
spring of spring-constant $k$. \ The damping constant $\tau$ is given by
$\tau=2e^{2}/(3mc^{3})$ and the random electric field $E_{x}$ is evaluated at
the center of the dipole. \ The random classical radiation drives the harmonic
oscillator into random oscillation with an average oscillator energy which is
the same as the random energy in the electromagnetic field at the same
frequency as that of the oscillator. \ Indeed this calculation was carried out
for random classical radiation by Max Planck in the last years of the 19th
century.\cite{Planck} \ However, Planck thought in terms of thermal radiation
at non-zero temperature, and did not consider the possibility of random
classical radiation at zero temperature. \ 

If one takes the limit as the charge $e$ becomes small (or indeed $e$ goes to
zero), the oscillator mechanical motion becomes uncoupled from the random
radiation. \ Sometimes one speaks of the mechanical oscillator as behaving
according to \textquotedblleft stochastic mechanics.\textquotedblright%
\ \ Indeed, if the random radiation takes on the Planck thermal spectrum
(including zero-point radiation) before taking the small-$e$ limit, then the
average values for position $x^{n}$ and for momentum $p^{n}$ for any power $n$
are identical between classical and quantum theory at every temperature
$T\geq0$. \ Also, the average value of the oscillator energy is the same
between classical and quantum theories. \ 

Both classical and quantum theories incorporate randomness, but the theories
are very different. \ The theories diverge when considering the fluctuations
of the energy or the average values of $x^{m}p^{n}$ involving products of
positions and momenta. \ Stochastic electrodynamics is a classical theory and
the harmonic oscillator (for $e>0)$ behaves as a random classical system; the
energy of the oscillator changes continuously as it absorbs energy from the
random radiation and radiates away energy according the the rules of Newton's
laws and Maxwell's equations.\cite{HB2} \ On the other hand, the quantum
oscillator behaves like a system which is never found in familiar classical
theory; at zero temperature, the quantum oscillator energy is absolutely fixed
at $\mathcal{E(}\omega)=\hbar\omega/2,$ with no fluctuations whatsoever, while
the position and momentum indeed fluctuate. \ At all temperatures (including
zero temperature) and for charge $e>0$, the classical oscillator is exchanging
energy continuously with the random radiation, both emitting radiation
continuously on acceleration and absorbing energy continuously from the random
radiation. \ At non-zero temperature, the quantum oscillator is exchanging
energy with the radiation field in only fixed amounts of energy, in
\textquotedblleft quanta\textquotedblright\ of magnitude $\hbar\omega.$ \ At
zero temperature, the quantum oscillator does not exchange energy at all; it
does not radiate away energy and it does not absorb energy. \ 

\subsection{Harmonic Oscillators Used to Describe Natural Phenomena}

Harmonic oscillators are used to describe many aspects of natural phenomena.
\ Thus solids are often modeled as lattices of molecules described as harmonic
oscillators; interacting molecules are often modeled by interacting harmonic
oscillators. \ Because of the agreement at equilibrium of the average energies
of the classical and quantum oscillators at all temperatures, there is
complete agreement between the results of classical and quantum calculations
involving charged harmonic oscillators. \ Thus the decrease of specific heats
at low temperatures can be accounted for in both the classical and quantum
theories. \ Also van der Waals forces between molecules modeled as harmonic
oscillators (both unretarded van der Waals forces and retarded forces) are
completely in agreement between classical and quantum systems. \ Furthermore,
the diamagnetic behavior of molecules modeled as three-dimensional harmonic
oscillator systems is in complete agreement between classical and quantum
theories. \ 

\subsection{Absence of Atomic Collapse}

In 1911 when Rutherford's scattering experiments with alpha particles
suggested that atoms contained a small positively charged nucleus, the
physicists of the period felt that they confronted the \textquotedblleft
problem of atomic collapse.\textquotedblright\ \ According to classical
electromagnetic theory, an accelerating charge will radiate away
electromagnetic energy. \ Thus an accelerating electron in a Coulomb orbit
around a small nucleus must be losing energy continuously through radiation.
\ Since the physicists of that period saw no source of radiation for the
electron, they anticipated that according to classical theory, the electron
must collapse into the nucleus in a very short time. \ In order for an atom to
be stable, some part of the classical description had to change. \ Bohr's
suggestion was that the rules of classical physics should be changed over to
new \textquotedblleft quantum\textquotedblright\ rules; the quantum rules
claimed that (contrary to classical electromagnetic theory) in certain
\textquotedblleft stationary states,\textquotedblright\ the accelerating
electrons did not radiate. \ The ad hoc rules prevented atomic collapse. \ 

What the physicists of 1911 did not consider, was the possibility of classical
zero-point radiation. \ If classical electromagnetic zero-point energy is
present, then there is a possibility that the electron will absorb energy from
the random zero-point radiation while losing energy as radiation while
accelerating. \ The situation would be entirely analogous to that for a
harmonic oscillator where (in the classical description) energy was being both
absorbed and emitted. \ Unfortunately, the classical calculations involving a
charged particle in a Coulomb potential are vastly more difficult than the
linear equations involving the harmonic oscillator. \ However, it is possible
(though difficult) to do numerical calculations. \ Numerical calculations for
a non-relativistic charged electron in a Coulomb potential were first carried
out by Cole and Zou\cite{CZ} in 2003, with results strongly suggesting that
classical zero-point radiation indeed provided the basis for understanding an
average behavior for the electron in a Coulomb potential; the probability
distribution in position for the classical electron in zero-point radiation
agreed fairly closely with the quantum-mechanical ground state distribution.
\ This successful calculation was challenged in 2015 by Nieuenhuisen and
Liska\cite{NL} who claimed that according to their more powerful calculations,
the electron indeed did not fall into the nucleus but rather was self-ionized,
because the electron picked up too much energy from the zero-point energy.
\ Both sets of calculations remove the atomic collapse problem which troubled
the physicists at the beginning of the 20th century. \ However, the
discrepancies between the two calculations indicate that more work on the
self-ionization aspect of this problem is needed.

\subsection{Importance of Relativity}

The physicists who introduced quantum theory at the beginning of the 20th
century missed two aspects of classical physics which today are considered
crucial for understanding natural phenomena. \ One aspect is the presence of
classical electromagnetic zero-point radiation; this aspect is explicitly
incorporated into stochastic electrodynamics. \ The second aspect is the
importance of special relativity. \ 

Relativity is important for stochastic electrodynamics because classical
electrodynamics is a relativistic theory. \ Indeed, the spectrum of classical
electromagnetic zero-point radiation is Lorentz invariant. \ It turns out that
there are two simple (at least approximately) relativistic systems. \ A
(relativistic) point charge in a Coulomb potential when considered within
classical electromagnetic theory is fully relativistic. \ A simple harmonic
oscillator system of small amplitude when considered within classical
electrodynamics is relativistic in the approximation of small amplitude, so
that the speed of the harmonic oscillator particle is small compared to the
speed of light $c.$

Relativity is important for understanding the hydrogen atom within stochastic
electrodynamics. \ It was pointed out in 1982 by Claverie and Soto\cite{CS}
\ that an analytic treatment of a nonrelativistic charged particle in a
Coulomb potential in the presence of zero-point radiation led to
self-ionization through the plunging particle orbits. \ However, the plunging
particle orbits are exactly the ones which are strongly modified by
relativity.\cite{B2004} \ Thus the conclusion of self-ionizing behavior based
upon nonrelativistic analysis is suspect. \ In addition to the situation for
the classical hydrogen atom, it turns out that relativistic behavior is
vitally important for understanding the blackbody problem within classical physics.

\subsection{Planck Spectrum of Blackbody Radiation}

It was in connection with Wien's suggestion for the blackbody spectrum that
Planck's constant $h$ was first introduced into physics. \ Subsequently,
Planck was able to extract the constant which we call \textquotedblleft
Boltzmann's constant,\textquotedblright\ $k_{B}=R/N_{A}$ (where $R$ is the gas
constant and $N_{A}$ is Avogadro's number), so that the thermal part of his
experimentally-based interpolation-guess for the blackbody spectrum
corresponded to an energy per normal mode $\mathcal{E(}\omega)=\hbar
\omega/[\exp(\hbar\omega/k_{B}T)-1].$ \ Although Planck turned to statistical
arguments in order to give a derivation of his interpolation-guess, other
physicists of the period tried to derive the blackbody radiation spectrum from
purely classical arguments. \ However, none of these physicists included the
idea of classical zero-point radiation, and so all arrived at the
Rayleigh-Jeans spectrum corresponding to an energy $\mathcal{E(}\omega
)=k_{B}T$ for each radiation normal mode. \ Indeed, Rayleigh and Jeans
considered classical statistical mechanics in deriving this radiation spectrum
for thermal radiation. \ 

The presence of classical electromagnetic zero-point radiation modifies all
the classical arguments advanced by the physicists in the first quarter of the
20th century. \ If zero-radiation is present, then traditional classical
statistical mechanics with its equipartition theorem is no longer valid. \ The
equipartition theorem suggests that an oscillator has zero energy at the
absolute zero of temperature; however, this result is contradicted within
stochastic electrodynamics since zero-point radiation gives a non-zero energy
to an oscillator at zero temperature. \ Significantly, several of the
arguments advanced by the physicists of the early 20th century can be easily
modified when zero-point radiation is introduced, and the modified
arguments\cite{HIst} lead directly to Planck's spectrum including zero-point
energy, corresponding to an energy per normal mode $\mathcal{E(}\omega
)=\hbar\omega/[\exp(\hbar\omega/k_{B}T)-1]+\hbar\omega/2=(\hbar\omega
/2)\coth[\hbar\omega/(2k_{B}T)].$ \ 

However, there remain a set of classical scattering calculations (using
nonrelativistic mechanical systems as scatterers) which lead to the
Rayleigh-Jeans spectrum. \ These non-relativistic scattering calculations are
indeed valid, but only at long wavelengths where relativity is not important
and where the Rayleigh-Jeans spectrum is an appropriate approximation.
\ Classical zero-point radiation is Lorentz-invariant, and it is unchanged
under scattering only if a relativistic scatterer is considered. \ Recently a
relativistic scattering calculation for zero-point radiation has been carried
out, and indeed detailed balance is found.\cite{Det} \ Only classical analyses
which are valid relativistic calculations can lead to the Planck spectrum with
zero-point radiation. \ 

Further evidence for the connection between relativity and the Planck spectrum
is found in calculations which are usually associated with general relativity.
\ The correlation functions found in a Rindler frame (an accelerating
relativistic frame) accelerating through zero-point radiation are associated
with the Planck spectrum. \ Indeed conformal transformations in a Rindler
frame can carry zero-point radiation into thermal radiation at non-zero
temperature.\cite{Rind}

\section{Phenomena Suggestively Connected to Stochastic Electrodynamics \ }

\subsection{Particle Diffraction}

It is often stated that the wave-like interference pattern found when
microscopic particles pass through two slits represents the essence of quantum
behavior. \ At present, this phenomenon has not been calculated within
stochastic electrodynamics. \ However, stochastic electrodynamics can provide
a suggestive description of the experimental observations. \ Whereas the
traditional description views the microscopic particles as moving in empty
space when passing through the slits, this is not the appropriate description
within stochastic electrodynamics. \ If zero-point radiation is present, all
particles will experience random forces due to the zero-point radiation.
\ Also, the slits through which the particles must pass will modify the
correlation functions associated with the zero-point radiation. \ Thus with
the stochastic electrodynamic description, the microscopic particles are
experiencing random forces due to zero-point radiation, and the correlation
functions for the zero-point radiation reflect the information about the slits
through which the microscopic particles will pass. \ Thus in the stochastic
electrodynamic view, it is reasonable that the pattern of the microscopic
particles after passing through the screen reflects the random forces
experience by the particles due to the zero-point radiation. \ Indeed, we know
that the average van der Waals forces between molecules and conducting walls
are exactly the same in classical and quantum theories, even for the
long-range van der Waals forces involving electromagnetic radiation. \ Perhaps
this classical-quantum agreement persists when dealing with the forces when
there is relative motion of a particle relative to the wall with slits.
\ Furthermore, at higher temperatures where the electromagnetic radiation
correlation functions should change according to the ideas of stochastic
electrodynamics, the particle interference pattern should change accordingly.
\ At present, the required calculations which can test the qualitative
descriptions arising in stochastic electrodynamics have not been carried out
for particles passing through slits. \ 

\subsection{Excited States and Spectral Lines}

The appearance of sharp spectral lines associated with transitions between
excited atomic states presents another basic aspect of nature for which
stochastic electrodynamics has only a qualitative explanation. \ The harmonic
oscillator is a system allowing easy calculations within both classical and
quantum theories. \ However, stochastic electrodynamics has no excited states
for the harmonic oscillator. \ At non-zero temperatures, the classical
oscillator has a continuous distribution of energies above the average
zero-point energy. \ In contrast, the quantum oscillator has discrete excited
states which are occupied with a Boltzmann probability distribution, giving
exactly the same average energy distribution as is found for the classical
oscillator in the Planck spectrum of random classical radiation. \ However,
the selection rules for quantum dipole transitions between the excited states
are such as to give radiation emission at only the fundamental frequency
$\omega$ of the harmonic oscillator. \ Indeed, in the dipole approximation, an
excited classical or quantum oscillator radiates energy at only the oscillator
frequency $\omega,$ and the decay rate is exactly the same between the
classical and quantum descriptions. \ 

For a harmonic oscillator, the appearance of the discrete quantum excited
states requires that the analysis goes beyond the dipole approximation over to
forbidden transitions, corresponding to an expansion of the electromagnetic
field as $E_{x}(x,t)=E_{x}(0,t)+x[\partial_{x^{\prime}}E(x^{\prime
},t)]_{x^{\prime}=0}+...$ \ But such an expansion beyond the dipole
approximation can also be taken for the treatment of the classical harmonic
oscillator in stochastic electrodynamics, and corresponds to the introduction
of parametric resonances for the oscillator. \ Indeed, Huang and
Batelaan\cite{HB} have shown that when a pulse of light is incident upon a
collection of classical or quantum oscillators, the absorption of the light is
resonant when the carrier frequency of the light corresponds to the harmonics
of the oscillator frequency $\omega,$ namely frequencies $n\omega\,,$
$n=1,2,3,...$\ \ Both the classical and the quantum descriptions absorb the
same amounts of radiation energy at the same frequencies. \ These sharp
absorption lines occur because of parametric resonance for the classical
oscillator in stochastic electrodynamics, and because of quantum excited
states for the quantum oscillator. \ This is a fascinating result and may
point the direction toward understanding more about atomic spectra in terms of
classical physics. \ At the moment, there is no analysis of radiation emission
for the classical hydrogen atom within stochastic electrodynamics, though Cole
and Zou\cite{CandZ} have found striking subharmonic resonances.

\subsection{Photon Behavior}

The idea of photons was introduced by Einstein in the early years of the 20th
century. \ In 1909, Einstein's analysis of the fluctuations in Planck's
spectrum of blackbody radiation claimed that there were two aspects for the
fluctuations. \ One aspect of the fluctuations agreed with classical wave
theory and the other aspect was associated with particle-like behavior with
particle energy $\hbar\omega.$ \ Crucially, Einstein's analysis did not
include the possibility of classical electromagnetic zero-point radiation.
\ Indeed, when classical electromagnetic zero-point radiation is introduced,
Einstein's analysis immediately gives a classical analysis for the Planck
spectrum including zero-point radiation.\cite{HIst} \ Furthermore, Einstein's
fluctuation analysis can be completely understood in terms of purely classical
physics where the particle-like \textquotedblleft photon\textquotedblright%
\ fluctuations are understood in terms of the interference between the thermal
contribution and the zero-point radiation contribution which is already
present when the thermal contribution is added. \ 

The photon behavior involving atomic radiative decay presents a more
complicated problem for stochastic electrodynamics. \ Thus a classical dipole
oscillator radiates away its energy in a continuous wave of radiation which
spreads out smoothly in all directions of space. \ In contrast, the
experimental data from x-ray emission suggest that the radiation is absorbed
in a specific direction and that the atom recoils in the opposite direction
from which the radiation was absorbed. \ This localized absorption and
associated recoil is currently described within a quantum photon model which
claims that the radiation from an atom is not emitted as a smooth classical
wave emerging in all directions but rather is emitted as a localized photon
moving in a specific direction. \ 

Within a qualitative analysis, the directed emission aspect remains a
possibility for stochastic electrodynamics. \ What the traditional quantum
explanation omits is the presence of random zero-point radiation which
interferes with the radiation which is emitted by the atom. \ It is the
interference which gives the possibility of directed behavior, just as
interference was crucial for a classical understanding of the blackbody
fluctuations discussed by Einstein. \ Indeed, a complete classical calculation
has been carried out for the much simpler situation of a pulsed current sheet
located in a standing wave of varying phase.\cite{inter} \ If the
surface-current pulse occurs in empty space, then the emitted radiation
spreads out symmetrically on both sides of the current sheet. \ However, in
the presence of a standing wave, the interference between the radiation
emitted by the surface-current pulse and the standing wave leads to radiation
energy which is directed to one side or the other while the emitting surface
experiences a recoil force in the opposite direction. \ This complete
classical calculation suggests a basis for understanding directed-emission
behavior in terms of the interference with the classical electromagnetic
zero-point radiation which must be present in the space according to
stochastic electrodynamics. \ 

\subsection{Superfluid Behavior}

In the quantum literature, it is suggested that the superfluidity of liquid
helium below the lambda point arises due to the high zero-point energy of the
helium atoms. \ Since stochastic electrodynamics includes classical zero-point
radiation which induces zero-point energy in all systems interacting with
electromagnetic radiation, we expect that stochastic electrodynamics might be
able to describe superfluidity from a classical viewpoint. \ Indeed there is
already suggestive work in this direction. \ Back in 1910, Einstein and Hopf
noted that a moving classical particle, which contained an internal classical
harmonic-oscillator interacting with random classical radiation, would
experience a random walk in velocity space; the random impulses are delivered
to the oscillator by the fluctuating radiation and damping is provided by the
average velocity-dependent force due to the motion of the harmonic oscillator
through the random radiation. \ Indeed, Einstein and Hopf used this model to
derive the Rayleigh-Jeans radiation spectrum for blackbody radiation. \ What
Einstein and Hopf did not anticipate was the possibility of classical
electromagnetic zero-point radiation. \ When classical zero-point radiation is
introduced (as in stochastic electrodynamics), then an equilibrium situation
at zero temperature requires that there must be an additional damping force
because zero-point radiation will provide random forces but does not give any
velocity-dependent damping. \ The additional damping is associated with the
radiation emission when the particle is accelerated at the walls of the
confining box. \ A reanalysis of the Einstein-Hopf model led to a derivation
of Planck's blackbody radiation spectrum including zero-point
radiation.\cite{B1969} \ At high temperatures $T$, the velocity-dependent
damping is dominant and the particle in the confining (one-dimensional) box
has an average kinetic energy $k_{B}T/2.$ \ However, as the temperature $T$
decreases toward absolute zero, the velocity-dependent damping force (which
depends upon the thermal part of the random radiation) decreases steadily
whereas the the random impulse due to both the thermal contribution and the
zero-point contribution goes over to a temperature-independent random impulse.
\ Zero-point radiation is Lorentz invariant and cannot give rise to
velocity-dependent forces. \ Thus at sufficiently low temperatures, the moving
particle experiences a random walk in velocity with only the accelerations at
the walls providing an energy loss which will lead to equilibrium. \ The
change from a velocity-dependent damping at high temperatures over to an
acceleration-dependent damping at the walls allows the possibility that the
particle behavior will be quite different at high and low temperatures.
\ Indeed, the expected low-temperature behavior has frictionless aspects which
are usually associated with superfluidity.\cite{Brownian}

\section{Connections Between Classical and Quantum Theories}

Traditional classical physics does not contain Planck's constant whereas
quantum physics does contain $\hbar.$ \ At the present time, all the textbooks
of modern physics present the dichotomy between classical and quantum physics
in terms of the presence of this constant. \ And some physicists would like to
keep the situation this way for pedagogical reasons. \ Within this traditional
distinction between the theories, there is a concern for the \textquotedblleft
infamous boundary\textquotedblright\ between the classical and quantum
theories which invoke differing theoretical concepts. \ 

The theory of stochastic electrodynamics upsets this facile distinction
between classical and quantum theories. \ Stochastic electrodynamics is a
purely classical theory which includes randomness and contains Planck's
constant. \ All the concepts of stochastic electrodynamics are
classically-recognized ideas, and the boundary between traditional classical
physics and stochastic electrodynamics involves simply how prominent a role is
played by the random classical zero-point radiation. \ 

How much of nature can be described by stochastic electrodynamics? \ The
answer to this question is still not known. \ In contrast to the enthusiastic
contributions of thousands of physicists to quantum theory, the predictions of
stochastic electrodynamics have been calculated by a tiny group of scientists.
\ Thus we cannot yet answer confidently as to how much of nature requires a
quantum description, or where the boundary lies between classical and quantum
physics. \ At present all that we can report with assurance is that stochastic
electrodynamics appears to be the closest classical theory to quantum theory,
and that stochastic electrodynamics can describe far more of nature at the
microscopic scale than can traditional classical physics.

\section{Acknowledgement}

This review was prepared in connection with the conference on stochastic
electrodynamics, SED2018-Boston July 18-20, organized by Professor Herman
Batelaan, Professor Anna Maria Cetto, and Professor Daniel Cole. \ I wish to
express my thanks to the organizers and participants for such a pleasant conference.

\end{document}